\documentclass{article}
 
\usepackage[onehalfspacing]{setspace}

\usepackage{mystyle}

\title{Nonhomothetic CES Preferences: \\
Closed-Form, Aggregation, and Microfoundation Results}

\title{Aggregation and Closed-Form Results\\
 for Nonhomothetic CES Preferences}

\author{Clement E. Bohr\thanks{Northwestern University. Email: \mbox{cebohr@u.northwestern.edu}} \and Mart\'i Mestieri\thanks{UPF, CREi, BSE and FRB of Chicago. Email: \mbox{marti.mestieri@gmail.com}} \and Emre Enes Yavuz\thanks{Northwestern University. \mbox{eey@u.northwestern.edu}}
}

\begin{document}

\maketitle

\begin{abstract}
\singlespacing \noindent
We provide four novel results for nonhomothetic Constant Elasticity of Substitution preferences \citep{h75}. First, we derive a closed-form representation of the expenditure function of nonhomothetic CES under relatively flexible distributional assumptions of demand and price distribution parameters. Second, we characterize aggregate demand from heterogeneous households in closed-form, assuming that household total expenditures follow an empirically plausible distribution. Third, we leverage these results to study the Euler equation arising from standard intertemporal consumption-saving problems featuring within-period nonhomothetic CES preferences. Finally, we show that nonhomothetic CES expenditure shares arise as the solution of a discrete-choice logit problem.
\end{abstract}

\section{Introduction}
Nonhomothetic Constant Elasticity of Substitution (CES) preferences were proposed by \cite{h75} and first applied to a general equilibrium setting by \cite{clm21}. Nonhomothetic CES preferences have proven useful since they can be integrated into standard applied general equilibrium models that rely on the assumption of constant price elasticity of preferences or production functions. They have been subsequently adopted in applied general equilibrium exercises in international trade, technology adoption, wage inequality, business cycles, and to assess the importance of nonhomotheticities for welfare measurement, e.g., among others,  \cite{matsuyama19}, \cite{zhang21}, \cite{l19}, \cite{cdm20}, \cite{jl22} and \cite{bb23}. 

Nonhomothetic CES preferences (or production functions) are defined through an implicit equation. This implicit formulation carries over, in general, to the expenditure function. Moreover, since these preferences do not belong to the Gorman class, they do not generally admit a representative consumer. 
This paper presents a particular case in which these results are overturned. First, we provide a set of sufficient parametric assumptions such that the expenditure function admits a closed form. This formulation allows for an arbitrary correlation between taste parameters and prices faced by a household. Moreover, it allows for a relatively rich distribution of goods with different income elasticities---a gamma distribution. The gamma distribution is flexible enough to generate a density function that is strictly declining or hump-shaped.\footnote{The gamma distribution is generically skewed, but it also converges to a normal distribution for an arbitrarily large "shape parameter," which we define in the next section.}

Second, building on the previous closed-form result, we show that under an empirically relevant distribution of household total expenditures, we can characterize the aggregate demand of each good in closed form. To obtain this result, we assume that households' total expenditures are distributed according to the Amoroso distribution. This distribution encompasses many well-known distributions as particular cases, e.g., log-normal or Pareto, and it has been shown to fit household income data remarkably well \citep{salem74}. Indeed it was developed and applied by Amoroso himself in 1925 to provide an accurate representation of the income distribution \citep{amoroso25}.

Third, building on the closed-form representation of the expenditure function, we explore the Euler equation arising from a standard intertemporal consumption/savings problem in which the intertemporal utility is CRRA. We provide a series of results depending on the CRRA coefficient. Finally, we also provide a derivation of nonhomothetic CES emerging from a discrete-choice logit model, analogous to that in \cite{adpt87} for the homothetic CES.\footnote{See also \cite{trottner19} for an independent derivation in the context of production functions. }

The rest of the note is structured as follows. Section \ref{sec:nh} briefly sketches the general definition of nonhomothetic CES. Section \ref{sec:gamma} provides the closed-form derivation of the expenditure function and Section \ref{sec:amoroso} provides the aggregation results across households with heterogeneous expenditure levels. We provide a description of the Euler equation under these particular cases in Section \ref{sec:euler}, and present the discrete-choice result in Section \ref{sec:logit}.

\section{Nonhomothetic CES Preferences}\label{sec:nh}

We define utility $U$ over a continuum of goods indexed by $i \in [0,1]$ implicitly through a nonhomothetic CES aggregator
\begin{equation}
    1 = \int_0^1\left(\Omega_i  U^{-\varepsilon_i} C_i \right)^{\frac{\rho-1}{\rho}} d i, \label{eq:nhces}
\end{equation}
where $C_i>0$ and  $\Omega_i>0$ denote the consumption and the taste parameter for good $i$, respectively, $\varepsilon_i \geq 0$ governs the expenditure elasticity for good $i$ and $\rho\in(0,1)\cup (1,\infty)$. In what follows, we assume $\varepsilon_i$ and $\Omega_i$ follow a joint distribution such that Equation \eqref{eq:nhces} yields a $U$ that is unique and quasi-concave in $\{C_i\}_{i\in [0,1]}$. \cite{h75} and \cite{clm21} provide sufficient conditions for this to be the case. They also show that the Hicksian demand for good $i$ implied by maximizing utility $U$ defined in Equation \eqref{eq:nhces} subject to the budget constraint $\int_0^1p_iC_i di=E$ is 
\begin{equation}
    C_i = \left( \frac{p_i}{E} \right)^{-\rho} \left[ \Omega_i U^{-\varepsilon_i} \right]^{\rho-1},\label{eq:ci}
\end{equation}
where $p_i$ is the price of good $i$ and the total expenditure function is,
\begin{equation}
    E = \left[\int_0^1 \left(\frac{p_i}{\Omega_i} U^{\varepsilon_i}\right)^{1-\rho} d i\right]^{\frac{1}{1-\rho}}. \label{tot_exp}
\end{equation}

\section{Closed-Form Solutions}\label{sec:gamma}

We first show that the mapping between total expenditures and utility given by the Equation \eqref{tot_exp} has a closed-form solution under the following two conditions on the joint distribution of $\varepsilon_i$ and $\Omega_i$:
\begin{assumption}\label{a1}
    The prices and taste parameters have a log-linear relationship with $\{\varepsilon_i\}_{i\in [0,1]}$,
    \begin{equation}
        \begin{gathered}
            \ln p_i = \xi_p \varepsilon_i + \nu_p,\\
            \ln \Omega_i = \xi_{\Omega} \varepsilon_i + \nu_{\Omega},\label{omega}
        \end{gathered}
    \end{equation}
    where $\nu_p$ and $\nu_\Omega$ can follow any distribution as long as the expectation $\mathbb{E} \left[ \left( \frac{e^{\nu_p}}{e^{\nu_\Omega}} \right)^{1-\rho} \right]$ exists. 
\end{assumption}

\begin{figure}[t]
\centering
\includegraphics[width=0.8\textwidth]{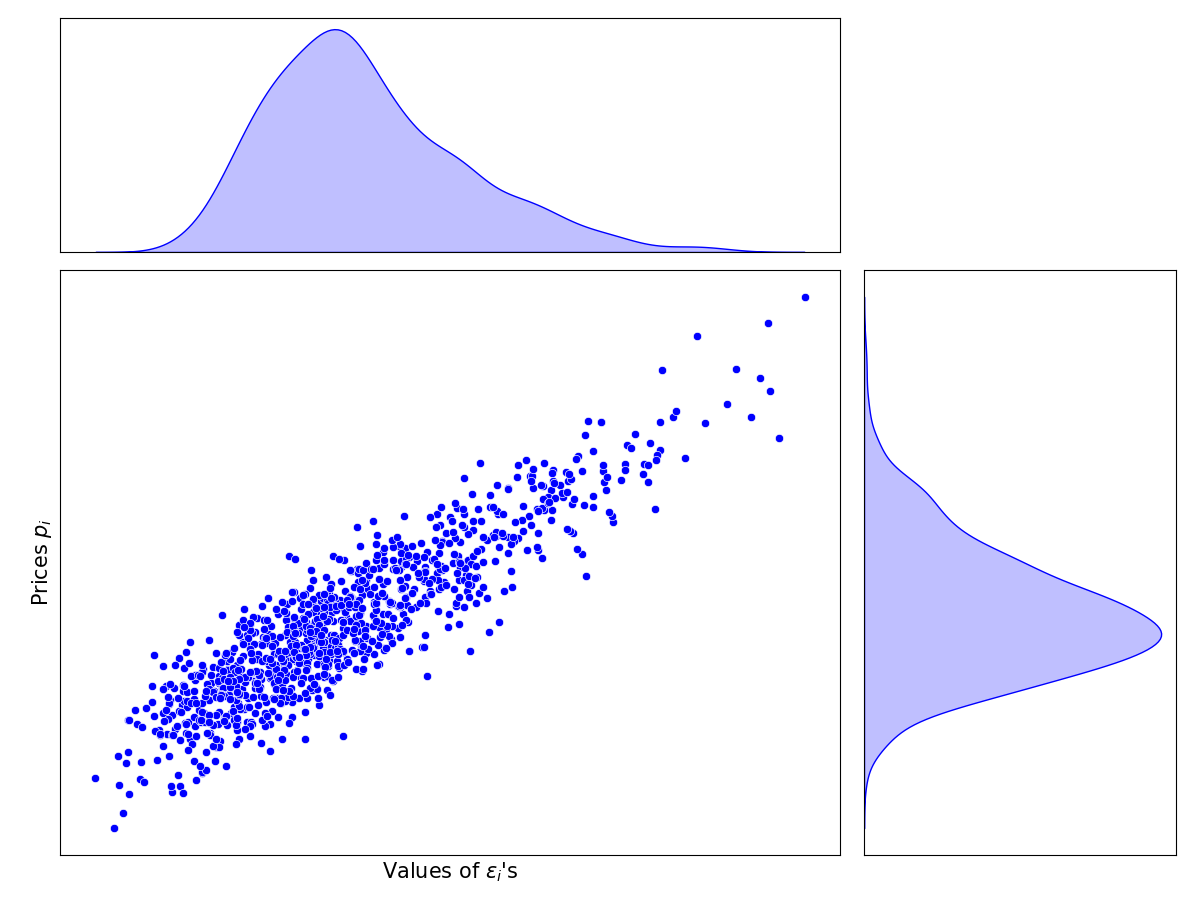}
\caption{An example of a joint distribution of $\varepsilon_i$'s and prices for which Assumption \ref{a1} and \ref{a2} hold.}
\end{figure}

The parameters $\xi_p$ and $\xi_\Omega$ in Equation \eqref{omega} govern the correlations between price, taste parameter, and expenditure elasticity. They can be an endogenous outcome of a general equilibrium model or exogenously specified.
\begin{assumption}\label{a2}
    $\{\varepsilon_i\}_{i\in [0,1]}$ are distributed following a gamma distribution,
    \begin{equation}
        \varepsilon_i \sim \text{Gamma}(\alpha, \beta), \label{eq:gamma}
    \end{equation}
    where $\alpha>0$ and $\beta>0$ are the shape and scale parameters of the gamma distribution. 
\end{assumption}
The density function is given by,
\begin{equation}
    f(\varepsilon) = \frac{\varepsilon^{\alpha-1} \exp \left[-\frac{\varepsilon}{\beta}\right]}{\beta^\alpha \Gamma(\alpha)}.
\end{equation}
As a reminder to the reader, when the shape parameter is less than or equal to one, $\alpha \leq 1$, the distribution has a tail shape, but for $\alpha > 1$, it becomes bell-shaped and a uni-modal distribution. For large values of $\alpha$, the distribution converges to the normal distribution, while the case $\alpha=1$ corresponds to the exponential distribution.
\begin{result}\label{res1}
   Suppose that Assumptions \ref{a1} and \ref{a2} hold. Then the mapping between utility and total expenditure is
\begin{equation}
\ln U=\frac{\Upsilon}{1-\rho}-\frac{\Psi}{1-\rho} E^{-\frac{1-\rho}{\alpha}} \label{eq: UE closed form}
\end{equation}
where,
\begin{equation}
\Upsilon=\frac{1}{\beta}-(1-\rho)\left(\xi_p-\xi_{\Omega}\right) \in \mathbb{R} \quad, \quad \Psi=\frac{\mathcal{M}^{\frac{1}{\alpha}}}{\beta} \in \mathbb{R}_{+} \quad , \quad \mathcal{M} \equiv \mathbb{E} \left[ \left( \frac{e^{\nu_p}}{e^{\nu_\Omega}} \right)^{1-\rho} \right]. \nonumber
\end{equation} 
Moreover, consumption choices $C_i$ in equation \eqref{eq:ci} are invariant to the scaling of $\beta$. Thus, we can normalize $\beta=1$ without loss of generality.
\end{result}
The closed-form result in Equation \eqref{eq: UE closed form} follows from direct integration of the expenditure function.\footnote{
After specifying the distribution of $p_i$ and $\Omega_i$, Equation \ref{tot_exp} becomes
\begin{align}
   E^{1-\rho} & =  \int_{\nu_\Omega} \int_{\nu_p} \int_0^\infty \left(  \frac{e^{\nu_p}}{e^{\nu_\Omega}} \right)^{1-\rho} \exp \left[(\xi_p - \xi_\Omega + \ln U ) \varepsilon ) \right]^{1-\rho} f(\varepsilon) f(\nu_p) f(\nu_\Omega)=\mathbb{E} \left[ \left( \frac{e^{\nu_p}}{e^{\nu_\Omega}} \right)^{1-\rho} \right]  \int_0^\infty \exp \left[(\xi_p - \xi_\Omega + \ln U ) \varepsilon ) \right]^{1-\rho} f(\varepsilon) \notag \\
    & = \mathbb{E} \left[ \left( \frac{e^{\nu_p}}{e^{\nu_\Omega}} \right)^{1-\rho} \right] \left( \frac{\tilde{\beta}}{\beta} \right)^\alpha \int_0^\infty  \frac{\varepsilon^{\alpha-1} \exp \left[-\frac{\varepsilon}{\tilde{\beta}}\right]}{\tilde{\beta}^\alpha \Gamma(\alpha)} = \mathbb{E} \left[ \left( \frac{e^{\nu_p}}{e^{\nu_\Omega}} \right)^{1-\rho} \right] \left( \frac{\tilde{\beta}}{\beta} \right)^\alpha,
\end{align}
where $\tilde{\beta} = \left( \frac{1}{\beta} - (1-\rho) (\xi_p - \xi_\Omega + \ln U) \right)^{-1}$ and the last step follows from the integral over a Gamma distribution density with parameters of $\alpha$ and $\tilde{\beta}$. This step requires $\Tilde{\beta} >0$, and it is possible always to ensure it.
} Using the closed-form mapping between the utility level and total expenditure in Equation \eqref{eq: UE closed form} and the demand Equation \eqref{eq:ci}, we find that the expenditure elasticity of good $i$ is
\begin{equation}
    \eta_i=\frac{\partial \ln C_i}{\partial \ln E}=\rho+(1-\rho) \mathcal{M}^{\frac{\beta}{\bar{\varepsilon}}} E^{\beta \frac{\sigma-1}{\bar{\varepsilon}}} \frac{\varepsilon_i}{\bar{\varepsilon}}. \label{eq: exp elas}
\end{equation}
where $\bar{\varepsilon}=\int_0^1 s_i \varepsilon_idi$ is the expenditure-share  weighted mean of $\varepsilon_i$ and $s_i$ denotes the expenditure share in good $i$.

Next, we show that Equation \eqref{eq: exp elas} is homogeneous of degree zero in $\beta$. This implies that expenditure elasticities and consumption choices are invariant to the choice of $\beta$. Suppose we scale $\beta$ proportionally by a factor $k>0$. Using the definition of utility, we find that this is equivalent to multiplying all $\varepsilon_i$'s in Equation \eqref{eq:nhces} with the same constant. More precisely, we have that if $\varepsilon_i$ is distributed gamma (i.e., Equation \eqref{eq:gamma} in condition 2 holds) then
\begin{equation}
    k \varepsilon_i \sim \text{Gamma}(\alpha, k \beta).
\end{equation}
Since the expenditure elasticity of good $i$ is pinned down by the value of $\varepsilon_i$ relative to its mean value $\overline{\varepsilon}_i \equiv \mathbb{E}[\varepsilon_i] = \alpha \beta $, the ratio of the two is independent of $k$. Therefore, $\beta$ can be normalized without losing generality. 

We conclude this section by noting that when the goods are complements, that is, $\rho < 1$, the goods with higher values of $\varepsilon_i$ are more expenditure elastic, and the expenditure elasticity \eqref{eq: exp elas} is always positive for all goods at any expenditure level. When the goods are substitutes, the relationship is reversed; higher $\varepsilon_i$ means less expenditure elastic goods.\footnote{Moreover, the expenditure elasticity becomes negative for a good $i$ when the expenditure level exceeds the expenditure level $E^*_i=\left(\frac{\rho-1}{\rho} \mathcal{M}^{\frac{\beta}{\bar{\varepsilon}}} \frac{\varepsilon_i}{\bar{\varepsilon}}\right)^{-\beta \frac{\bar{\varepsilon}}{(\sigma-1)}}$. As the expenditure level increases, consumption increases relatively more on goods with a high expenditure elasticity (low $\varepsilon_i$). Since the gamma distribution is defined on the positive real line, $\varepsilon \geq 0$, eventually, the expenditure share of goods with $\varepsilon > 0$ has to decline.}


\section{Demand Aggregation of Households with Heterogeneous Total Expenditure Levels}\label{sec:amoroso}
We consider an economy populated by households with heterogeneous expenditures satisfying Result \ref{res1}, so that they have an expenditure function given by Equation \eqref{eq: UE closed form}. In this case, we show that if total expenditure levels follow an Amoroso distribution, the aggregate demand has a closed-form solution. The Amoroso distribution is named after the Italian economist Luigi Amoroso who used it to study income distributions \cite{amoroso25}. A central motivation for Amoroso was to address the counterfactual feature of the Pareto distribution of having a mode at the lower bound of its support. This distribution appears to provide a reasonably good representation of total household expenditures and income \citep{salem74}.

Consider household $h$ with total expenditure $E_h$ given by Equation \eqref{eq: UE closed form}. In this case, the expenditure share in good $i$
is \begin{equation}
   s_{ih} = \exp(\varepsilon_i \Upsilon) \left( \frac{\Omega_i}{p_i} \right)^{\rho-1} E_h^{\rho-1} \exp \left( -\varepsilon_i \Psi E_h^{\frac{\rho-1}{\alpha} } \right).
\end{equation}
We assume the expenditure distribution follows an Amoroso distribution, whose probability density function is given by,
\begin{equation}
    f_{E_h}(x \mid l, k, m, n)=\frac{1}{\Gamma(m)}\left|\frac{n}{k}\right|\left(\frac{x-l}{k}\right)^{m n-1} \exp \left\{-\left(\frac{x-l}{k}\right)^n\right\}.\label{eq:amo}
\end{equation}
with $m>0$ and $k,l,n \in\mathbbm{R}$. We shall focus on the economically relevant case $k>0$. In this case, the support of the distribution is $x \geq l$. The Amoroso distribution has two shape parameters, $m$ and $n$, a scale parameter $k$, and a location parameter $l$. Since the units of measurement of expenditures are a free degree of freedom, we focus on the particular normalization of $l=0$ without loss of generality. 
 It is readily verified by direct substitution that the Amoroso distribution in Equation \eqref{eq:amo} has as particular cases many familiar distributions such as the exponential, gamma, Fr\'echet, Weibull distributions, and as limiting cases as $n\rightarrow 0$, the lognormal and Pareto distributions (depending on what we assume about $m$).\footnote{
Over fifty distinct distributions occur as special cases or limiting forms of the Amoroso distribution; see \cite{amoroso}.} 
\begin{assumption}\label{a3}
    The second shape parameter $n$ of the Amoroso distribution \eqref{eq:amo} satisfies $n=\frac{\rho-1}{\alpha}$. The location parameter is normalized to zero, $l=0$ without loss of generality.
\end{assumption}

\begin{figure}[t]
\includegraphics[width=1.1\textwidth]{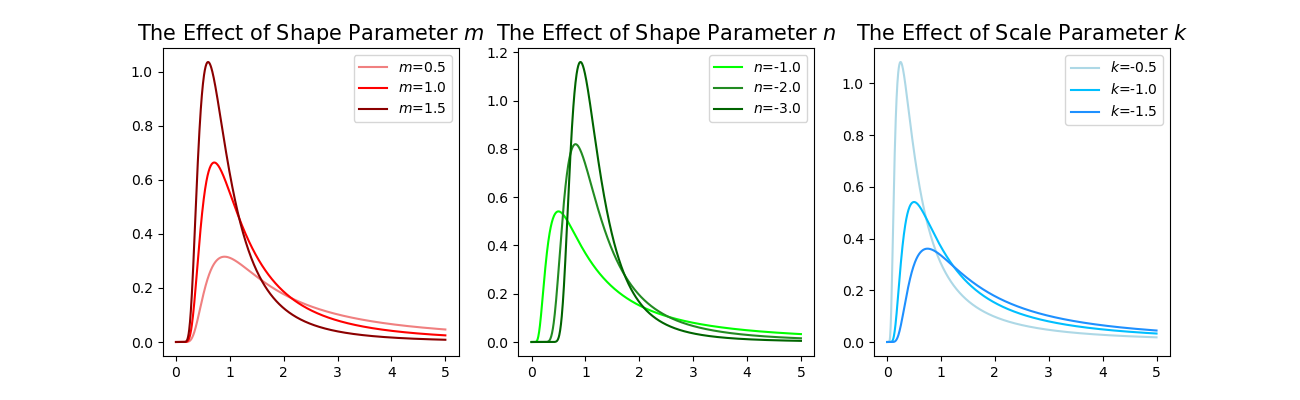}
\caption{Amoroso distribution with different shape parameters.}
\label{fig:amoroso}
\end{figure}

This assumption links the second shape parameter of the Amoroso distribution to the elasticity of substitution parameter $\rho$ and one of the parameters of the Gamma distribution of $\varepsilon_i$ appearing in Assumption \ref{a2}. So, when this assumption holds, we have only two degrees of freedom left in the Amoroso distribution, $k$ and $m$. Fortunately, varying these two parameters yields a wide range of potential shapes that can be bell-shaped or monotonically decreasing (e.g., Weibull, Fr\'echet or Pareto). Figure \ref{fig:amoroso} also shows that the changes in the distribution induced by the normalized parameter $n$ are qualitatively similar to those of $m$ and $k$. 
\begin{result}\label{res2}
    Consider an economy populated by a continuum of agents indexed by $h$, whose total expenditure, $E_h$, is distributed Amoroso (Equation \ref{eq:amo}). If Assumptions \ref{a1}, \ref{a2} and \ref{a3} hold, total expenditure shares of good $i$ in the population are 
\begin{equation}\label{eq:aggamo}
    s_i = \int s_{ih} f_{E_h}(E_h) dE_h = \exp(\varepsilon_i \Upsilon) \left( \frac{\Omega_i}{p_i} \right)^{\rho-1} \frac{\Gamma(m+\alpha)}{\Gamma(m)} \frac{k^{\rho-1}}{\left[ 1 + \varepsilon_i \Psi k^{\frac{\rho-1}{\alpha}} \right]^{m+\alpha} }
\end{equation} 
The expenditure share in terms of the average household expenditure,  $\bar E_h=k\frac{\Gamma\left(m+\frac{\alpha}{\rho-1}\right)}{\Gamma(m)}$, is 
\begin{equation}
     s_i = \int s_{ih} f_{E_h}(E_h) dE_h = \exp(\varepsilon_i \Upsilon) \left( \frac{\Omega_i}{p_i} \right)^{\rho-1} \frac{\Gamma(m+\alpha)}{\Gamma(m+\frac{\alpha}{\rho-1})} \frac{k^{\rho-2}}{\left[ 1 + \varepsilon_i \Psi k^{\frac{\rho-1}{\alpha}} \right]^{m+\alpha} } \bar E_h \label{eq:aggexact}
\end{equation}
which approximately corresponds to\footnote{The result follows from the approximation of the ratio of gamma functions $\Gamma(m+\alpha/(\rho-1))/\Gamma(m)\approx m^\alpha/(\rho-1)$, which implies $\Gamma(m+\alpha/(\rho-1))/\Gamma(m)\approx \left(\Gamma(m+\alpha)/\Gamma(m)\right)^{1/(\rho-1)}$. This approximation is better for $m\gg \alpha/(\rho-1)$, see \cite{gammaratio}.}
\begin{equation}\label{eq:aggamo}
    s_i \approx\exp(\varepsilon_i \Upsilon) \left( \frac{\Omega_i}{p_i} \right)^{\rho-1}  \frac{\bar E_h^{\rho-1}}{\left[ 1 + \varepsilon_i \Psi k^{\frac{\rho-1}{\alpha}} \right]^{m+\alpha} }.
\end{equation} 
\end{result}
Equations \eqref{eq:aggexact} and \eqref{eq:aggamo} show that the distribution of expenditures affects aggregate sectoral consumption---not surprisingly, since nonhomothetic preferences are not Gorman-aggregable. Perhaps interestingly, Equation \eqref{eq:aggexact} implies that aggregate consumption is log separable in the consumption choice of the household with average expenditures and another term that captures the importance of inequality in the form of higher moments of the expenditure distribution.

\section{Intertemporal Substitution and Euler Equation}\label{sec:euler}
Nonhomothetic CES preferences also affect the consumption/savings decision of households. The intertemporal substitution is affected because households account for the change in their consumption basket due to changes in their utility level when they consider how to substitute expenditure between time periods. In this note, we focus on the case in which the per-period flow utility is given by a CRRA function,
\begin{equation}
v(U)=\frac{U^{1-\theta}-1}{1-\theta}
\end{equation}
where $\theta$ controls the intertemporal elasticity of substitution. The household maximizes the discounted sum of flow utility,

\begin{equation}
    \max_{\{E_t, A_t\}_{t=0}^\infty} \sum_{t=0}^\infty \beta^t v(U) \label{eq: intertemporal max}
\end{equation}
subject to within-period preferences being the nonhomothetic CES preferences defined in Section \ref{sec:nh} and the intertemporal budget constraint
\begin{equation}
    E_t + A_{t+1} \leq (1+r_t) A_t + Y_t,
\end{equation}
where $A_t$ is a risk-free asset in zero net supply ($A_0 = 0$) and $Y_t$ is household income.

\begin{result}
 \begin{enumerate}
        \item[(i)] The Euler equation is given by,

        \begin{equation}
            \left( \frac{E_{t+1}}{E_t} \right)^\theta = \beta (1+r_t) \left( \frac{\overline{\varepsilon}_{t+1}}{\overline{\varepsilon}_t} \right)^{-1} \left( \frac{P_{t+1}}{P_t}\right)^{\theta-1} 
        \end{equation}
       where $P_t$ is the ideal price index defined as $\frac{E_t}{U_t}$ and $\overline{\varepsilon}_t$ is a weighted average $\varepsilon_i$ at time $t$, given by $\overline{\varepsilon}_t = \int_0^1 s_i\varepsilon_i  di$ with $s_i = \frac{p_{it} C_{it}}{E_t}$ being expenditure share of good $i$.

       \item[(ii)] If $p_i$ and $\Omega_i$ follow distributions with Assumption \ref{a1} holding, and hence with a closed form mapping between $U_t$ and $E_t$ given by Equation \eqref{eq: UE closed form}, then we have $\overline{\varepsilon}_t = \frac{\alpha}{\Psi} E_t^{\frac{1-\rho}{\alpha}}$. In addition, if we normalize the ideal price index to one, $P_t=1$, for all $t$  and assume $\theta+\frac{1-\rho}{\alpha} \geq 0$, then the Euler equation becomes,\footnote{
 Without normalizing the ideal price index, the Euler equation is given by
\begin{equation}
                \left( \frac{E_{t+1}}{E_t} \right)^{1 + \frac{1-\rho}{\alpha}} = \beta (1+r_t) \exp \left[ \frac{(\theta-1)\Psi}{1-\rho} \left( E_t^{-\frac{1-\rho}{\alpha}} - E_{t+1}^{-\frac{1-\rho}{\alpha}} \right) \right].
\end{equation}

 }
        \begin{equation}
            \left( \frac{E_{t+1}}{E_t} \right)^{\theta + \frac{1-\rho}{\alpha}} = \beta (1+r_t).
        \end{equation}

        \item[(iii)] If there are households with identical preferences and their total expenditures distributed Amoroso at time $t$, $E_{ht} \sim Amoroso(l,k,m,n)$ and the previous result holds, then the expenditure distribution at time $t+1$ is also Amoroso, $E_{ht+1} \sim Amoroso(l,A \times k,m,n)$ where $A = \left[ \beta (1+r_{t+1}) \right]^\frac{\alpha}{\alpha \theta + 1 - \rho}$.
        
    \end{enumerate}
\end{result}

The solution of the intertemporal optimization defined in the Equation \eqref{eq: intertemporal max} is given by 
\begin{equation}
\frac{ v^\prime (U_t) \frac{\partial U_t}{\partial E_t} }{ v^\prime(U_{t+1})   \frac{\partial U_{t+1}}{\partial E_{t+1}} } = \beta (1+r_{t+1} ).
\end{equation}
The household arranges its expenditure between periods so that the ratio of flow utility $v(U_t)$ that one additional dollar can buy equalizes to the discounted rate of return. The curvature driven by CRRA preferences is standard $v^\prime(U_t) = U_t^{-\theta}$. However, $\frac{\partial U_t}{ \partial E_t}$ is not constant due to preferences being nonhomothetic. In particular,
\begin{equation}
\frac{\partial U_t}{\partial E_t} =  \frac{ \overline{\varepsilon}_t^{-1} }{P_t}.
\label{eq:derivelas}
\end{equation}
By Walras' law, we can normalize $P_t=1$. Then, the derivative in \eqref{eq:derivelas} is only determined by the average $\varepsilon_i$ weighted by the expenditure shares. If the goods are complements, $\rho<1$, then the higher $\varepsilon_i$ means more expenditure elastic goods, so $\overline{\varepsilon}_t$ can be thought as an average luxury index of the consumption basket at a given time. Therefore, in this case, as the household gets richer and spends relatively more on more luxurious goods, one more dollar buys less utility. If Assumption \ref{a1} holds, $\overline{\varepsilon}_t \propto  E_t^{\frac{1-\rho}{\alpha}}$. Therefore the nonhomotheticity acts as if we had some additional curvature in intertemporal preferences (or lesser curvature if the goods are substitutes, $\rho>1$).

\section{Logit Microfoundation of Preferences}\label{sec:logit}
In this last section, we show that it is possible to derive the expenditure shares of the nonhomothetic CES as the probability choice of a logit model analogously to \cite{adpt87}. This provides an alternative aggregation result for nonhomothetic CES. Assume an economy populated by a mass of households with linear preferences over goods indexed by $i\in[0,1]$. Households optimally consume a single good according to their indirect utility function,

\begin{equation}
V_i = \ln \Omega_{i}+(1-\varepsilon_i) \ln \frac{E}{p}-\ln \frac{p_{i}}{p}+\mu \nu_{i}\label{eq:indu}
\end{equation}
where $\frac{E}{p} = U$ denotes the level of "real expenditures" according to the optimal utility-based price-index $p$, $\nu_{i}$ is an unobservable preference heterogeneity shock following a standard Gumbel distribution,\footnote{That is, with cumulative distribution is $\exp(\exp(-\nu))$.} and $\mu$ is a parameter governing the degree of preference heterogeneity. Lastly, $\Omega_{i}$ captures observable preference heterogeneity or tastes. It is well known that the probability of a household choosing to consume good $i$ takes a logit form. Under this utility specification, it becomes
\begin{equation}
\operatorname{Pr} ( i )=\frac{\left(\frac{E}{p}\right)^{\frac{1-\varepsilon_i}{\mu}}\left(\Omega_{i}^{-1} \frac{p_{i}}{p}\right)^{-\frac{1}{\mu}}}{ \int \left(\frac{E}{p}\right)^{\frac{1-\varepsilon_i}{\mu}}\left(\Omega_{i}^{-1} \frac{p_{i}}{p}\right)^{-\frac{1}{\mu}} d i }.
\end{equation}
We set $\mu = - \frac{1}{1-\rho}$ so that the denominator equals one by the definition of the utility. Furthermore, if all households have the same expenditure level, $E$, then the demand share of good $i$ is given by the probability above,
\begin{equation}
s_i=\frac{p_{i} C_{i}}{E}=\left(\Omega_{\varepsilon i}^{-1} \frac{p_{i}}{p}\right)^{1-\rho}\left(\frac{E}{p}\right)^{(\varepsilon_i-1)(1-\rho)},\label{eq:logitnhces}
\end{equation}
which is the same as the demand share implied by the nonhomothetic CES aggregator if we choose $p= \left[\int\left(\frac{p_i}{\Omega_i} U^{\varepsilon_i-1}\right)^{1-\rho} d i\right]^{\frac{1}{1-\rho}}$. We have shown thus the following result. 
\begin{result}
    Suppose that indirect utility is given by Equation \eqref{eq:indu} with $ E=\left[\int\left(\frac{p_i}{\Omega_i} U^{\varepsilon_i}\right)^{1-\rho} d i\right]^{\frac{1}{1-\rho}}$. Consider a population of mass 1 and expenditure $E$ in which individual $i$ maximizes $V_i$ given an independent draw of $v_i$. Aggregate expenditure shares in the economy coincide with those of a consumer with nonhomothetic CES yielding expenditure shares as in Equation \eqref{eq:logitnhces}.  
\end{result}

\section{Conclusion}
This note has presented sufficient conditions for a closed-form representation of the expenditure function of nonhomothetic CES and aggregation of demand of households with heterogeneous levels of total expenditure. We also analyzed the resulting Euler equation under this parameterization of nonhomothetic CES and provided a link with a discrete-choice model. We believe that these results may be of use in applications that rely on nonhomothetic CES formulation of preferences or production. Finally, we note that an analogous exercise can be performed to heterothetic Cobb-Douglas preferences \citep{bmrn23}.

\bibliography{biblio}

\end{document}